\magnification=1210
\baselineskip=15pt
\def\medskip{\vskip .2in}

\def\ds{\displaystyle}
\pageno=1
\footline{\hss\folio\hss}
\vskip2.5pc
\centerline{\bf  Constant of Motion for several one-dimensional systems and outlining}
\centerline{\bf  the problem associated with getting their Hamiltonians}
\vskip2pc
\centerline{ G. L\'opez, L.A. Barrera, Y. Garibo, H. Hern\'andez, J.C. Salazar, and C.A. Vargas
}
\vskip1pc
\centerline{Departamento de F\'isica de la Universidad de Guadalajara}
\centerline{Apartado Postal 4-137}
\centerline{44410~Guadalajara, Jalisco, M\'exico}
\vskip5pc
\centerline{Nov 2003} 
\vskip1pc
\centerline{PACS~~03.20.+i~~~03.65.Ca~~~03.30.+p~~~03.20.+i}
\centerline{keywords:~~ constant of motion}

\vskip4pc
\centerline{ABSTRACT}
\vskip2pc
The constants of motion of the following systems are deduced: a relativistic particle  with
linear dissipation, a no-relativistic
particle with a time explicitly depending force, a no-relativistic particle with a constant
force and time depending mass, and a relativistic particle under a conservative force with
position depending mass. The problem of getting the Hamiltonian for these systems
is determined by getting the velocity as an explicit function of position and generalized
linear momentum, and this problem can be solved a first approximation for the first above
system.

\vfil\eject
\leftline{\bf I. Introduction.}
\vskip0.5pc
The constant of motion of a dynamical system, which has an equivalent interpretation of the
energy of the system, has received attention lately due to the study of dissipative system
[1],  due to some problems with the Hamiltonian formalism [2], and due to the possibility of
making a quantum mechanics formulation based on the constant of motion concept [3]. The
constant of motion concept, besides its obvious usefulness in classical mechanics, can have
great deal of importance in quantum mechanics and statistical physics for system without well
defined Hamiltonian [3]. In particular, when one studies relativistic systems with
no-conservative forces or systems with time depending mass or systems with position depending
mass, the concept of constant of motion appears more naturally than the concept of
Hamiltonian. In this paper we analyze four dynamical systems and find their constant of
motion. These constants of motion are selected such that when some interaction is neglected,
they are reduced to the usual concept of energy. The Hamiltonian associated to the system is
deduced whenever is possible to do that. The paper is organized as follows: we study first a
relativistic system with linear dissipation and with a constant external force. For this
system, the constant of motion is given in general, and the Hamiltonian is obtained for weak
dissipation only. Then, we study a no-relativistic system with an external time explicitly
depending force, where only the constant of motion is given. 
In the same way, we find a constant of motion for a no-relativistic system with a constant
force and with a time depending mass.  Similarly and finally, we obtained the constant of
motion of a relativistic system with position depending mass and a force proportional to this
mass.
\vfil\eject
\leftline{\bf II. Constant of motion of a relativistic particle with linear dissipation.}
\vskip0.5pc
The motion of a relativistic particle with rest mass "$m$" and under a constant force "$f$" and
a linear dissipation is described by the equation
$${\ds d\over\ds dt}\left({mv\over\sqrt{1-v^2/c^2}}\right)=f-\alpha v\ ,\eqno(1)$$
where $v$ is the velocity of the particle, $c$ is the speed of light, and $\alpha$ is the
parameter which characterizes the dissipative linear force. Eq. (1) can be written as the
following autonomous dynamical system
$$\eqalign{
{\ds dx\over\ds dt}&=v\ ,\cr
{\ds dv\over\ds dt}&={\ds f\over\ds m}(1-\beta v)(1-v^2/c^2)^{3/2}\ ,\cr}\eqno(2)$$
where $\beta$ has been defined as $\beta=\alpha/f$, and $x$ is the position of the particle. A
constant of motion of this system is a function $K_{\beta}(x,v)$ [4] which satisfies the
equation 
$$v~{\partial K_{\beta}\over\partial x}+{f\over m}(1-\beta v)(1-v^2/c^2)^{3/2}{\partial
K_{\beta}\over\partial v}=0\ .\eqno(3)$$
The solution of Eq. (3), such that for $\beta$ equal to zero one gets the usual expression
for the relativistic energy,
$$\lim_{\beta \to 0}K_{\beta}={\ds mc^2\over\sqrt{1-v^2/c^2}}-fx-mc^2\eqno(4)$$ 
(the last term, $mc^2$, was added to obtain the right no-relativistic energy expression for $c$
going to infinity), is given by
$$K_{\beta}=-fx-mc^2+{\ds mc^2\over\ds\sqrt{1-{v^2\over c^2}}}\cases{
{\ds 1+\beta v\over\ds 1-\beta^2c^2}+
{\ds \beta c\sqrt{1-v^2/c^2}\over\ds (\beta^2c^2-1)^{3/2}}\ln{A_{\beta}(v)}& if
$\beta>1/c$\cr\cr {\ds 1+\beta v\over\ds 1-\beta^2c^2}+
{\ds \beta c\sqrt{1-v^2/c^2}\over\ds (1-\beta^2c^2)^{3/2}}\arctan{B_{\beta}(v)}& if
$\beta<1/c$\cr\cr 
{1\over 3}\left[{\ds v\over\ds c}-{\ds 1\over\ds 1-v/c}\right]&if $\beta=1/c$\cr}\eqno(5a)$$
where the functions $A_{\beta}(v)$ and $B_{\beta}(v)$ are defined as
$$A_{\beta}(v)={\ds 2(\beta^2c^2-\beta v)+2\beta c\sqrt{\beta^2c^2-1}\sqrt{1-v^2/c^2}\over\ds
1-\beta v}\eqno(5b)$$
and
$$B_{\beta}(v)={\ds \beta^2c^2-\beta v\over\ds \beta c\sqrt{1-\beta^2c^2}\sqrt{1-{v^2\over
c^2}}}\ .\eqno(5c)$$
At first order on the dissipation parameter, the constant of motion can be written as
$$K={\ds mc^2\over\ds\sqrt{1-{v^2\over c^2}}}-fx-mc^2+\beta mc^3\left[
{\ds v/c\over\ds\sqrt{1-{v^2\over c^2}}}-\arctan\left({v/c\over\sqrt{1-{v^2\over
c^2}}}\right)\right]\ .\eqno(6)$$
Now, using the known expression relating the constant of motion and the Lagrangian [5],
$$L=v\int{\ds K(x,v)~dv\over \ds v^2}\ ,\eqno(7)$$
this Lagrangian is calculated inserting (5a) in (7), bringing about the expression
$$
L_{\beta}=fx+mc^2+\cases{
{ mc^2\sqrt{1-{v^2\over c^2}}\over\beta^2c^2-1}+
{\beta c^2 vm \over\beta^2c^2-1}\ln\left[{2(1+\sqrt{1-v^2/c^2}~)\over\beta v}\right]+ 
{mc^2G_{\beta}(v)\over 4(\beta^2c^2-1)}&if $\beta>1/c$\cr\cr
{ mc^2\sqrt{1-{v^2\over c^2}}\over\beta^2c^2-1}+
{\beta c^2 vm\over\beta^2c^2-1}\ln\left[{2(1+\sqrt{1-v^2/c^2}~)\over\beta v}\right]+
{mc^3\beta Q_{\beta}(v)\over (1-\beta^2c^2)^2}&if $\beta<1/c$\cr\cr
{ mc^2\over 3\sqrt{1-v^2/c^2}}\left[1-{v\over c}-{2v^2\over c^2}+{v\over c}
\sqrt{1-{v^2\over c^2}}~R_{\beta}(v)\right]-{mcvR_{\beta}(v)\over 3}&if $\beta=1/c$
\cr}$$
$$\eqno(8)$$
where the functions $G_{\beta}$, $Q_{\beta}$ and $R_{\beta}$ are given in the appendix. For
week dissipation, one can use (6) in (7) to get
$$L=-mc^2\sqrt{1-{v^2\over c^2}}+fx+mc^2+\beta mc^3\arctan\left({v/c\over\sqrt{1-v^2/c^2}}
\right)
\ .\eqno(9)$$
The generalized linear momentum, $p=\partial L/\partial v$, can be calculated using (8),
bringing about the expression
$$\eqalign{
p_{\beta}=\cases{
{-mv\over(\beta^2c^2-1)\sqrt{1-{v^2\over c^2}}}+{\beta c^2m\over\beta^2c^2-1}
\ln\left[{2(1+\sqrt{1-{v^2\over c^2}}~)\over\beta v}\right]-
{\beta mc^2\over(\beta^2c^2-1)\sqrt{1-{v^2\over c^2}}}+A_{\beta}^{(1)}&if $\beta>1/c$\cr\cr
{-mv\over(\beta^2c^2-1)\sqrt{1-{v^2\over c^2}}}+{\beta c^2m\over\beta^2c^2-1}
\ln\left[{2(1+\sqrt{1-{v^2\over c^2}}~)\over\beta v}\right]-
{\beta mc^2\over(\beta^2c^2-1)\sqrt{1-{v^2\over c^2}}}+A_{\beta}^{(2)}&if $\beta<1/c$\cr\cr
{mc\over 3(1-{v\over c})\sqrt{1-{v^2\over c^2}}}\left({2v^2\over c^2}-{2v\over c}-1\right)
&if $\beta=1/c$\cr}}$$
$$\eqno(10)$$
where the functions $A_{\beta}^{(1)}$ and $A_{\beta}^{(2)}$ are given in the appendix. For
weak dissipation, expression (9), the generalized linear momentum is given by
$$p={mv+\beta mc^2\over\sqrt{1-{v^2\over c^2}}}\ .\eqno(11)$$
As one can see from (10), it is not possible to express $v$ explicitly as a function of
$p_{\beta}$. Therefore, it is not possible to know explicitly the Hamiltonian of the system.
However, at first order on the dissipation parameter, relation (11), one can get the following
expression
$$v={-\beta m^2c^2+p\sqrt{p^2/c^2+m^2-\beta^2m^2c^2}\over p^2/c^2+m^2}\ .\eqno(12)$$
So, the Hamiltonian for this weak dissipation case can be written as
$$H={mc^2(p^2/c^2+m^2)\over g_{_\beta}(p)}-fx-mc^2+\beta mc^3\Delta_{\beta}(p)\ ,\eqno(13a)$$
where $g_{_\beta}(p)$ and $\Delta_{\beta}(p)$ are functions defined as
$$g_{_\beta}=\sqrt{({p^2\over c^2}+m^2)^2-\beta^2m^2c^2+2\beta
m^2p\sqrt{{p^2\over c^2}+m^2-\beta^2m^2c^2} -{p^2\over
c^2}({p^2\over c^2}+m^2-\beta^2m^2c^2)}\eqno(13b)$$
and
$$\Delta_{\beta}={-\beta m^2c^2+p\sqrt{{p^2\over c^2}+m^2-\beta^2m^2c^2}\over
cg_{_\beta}(p)}-
\arctan\left[{-\beta m^2c^2+p\sqrt{{p^2\over c^2}+m^2-\beta^2m^2c^2}\over
cg_{_\beta}(p)}\right]
\eqno(13c)$$
Note that the function $g_{_\beta}$ has the following limit $\lim_{\beta\to
0}g_{_\beta}(p)=m\sqrt{p^2/c^2+m^2}$. Thus, (13a) has the usual Hamiltonian expression as
$\beta$ goes to zero.
\vskip1pc
\leftline{\bf II. Constant of motion for a time depending force.}
\vskip0.5pc
The motion of a no-relativistic particle of mass $m$ under a time depending force, $f(t)$, can
be written as the following non-autonomous dynamical system
$${dx\over dt}=v\ ,\eqno(14a)$$
$${dv\over dt}=f(t)/m\ .\eqno(14b)$$
A constant of motion for this system is a function $K(x,v,t)$ such that satisfies the
following equation [4]
$$v{\partial K\over\partial x}+{f(t)\over m}{\partial K\over\partial v}+{\partial
K\over\partial t}=0\ .\eqno(15)$$
Solving this equation by the characteristics method [6], one gets the general solution given by
$$K(x,v,t)=G(C_1,C_2)\ ,\eqno(16)$$
where $G$ is an arbitrary function of the characteristics $C_1$ and $C_2$ which has the
following expressions
$$C_1=v-{1\over m}\int f(t)~dt\ ,\eqno(17a)$$
and
$$C_2=x-vt+{t\over m}\int f(t)~dt-{1\over m}\int\left(\int^tf(s)~ds\right)~dt\ .\eqno(17b)$$
Let us choose $f(t)$ of the form
$$f(t)=f_o[1+\epsilon g(t)]\ ,\eqno(18)$$
where $g(t)$ is an arbitrary function, and $\epsilon$ and $f_o$ are parameters. Note that
$\lim_{\epsilon\to 0}f(t)=f_o$, and in this limit, the usual constant of motion is the energy,
$K_o=\lim_{\epsilon\to 0}K=mv^2/2-f_ox$. In order to get this energy expression from our
characteristics within this limit, one needs in (16) the following functionality
$\lim_{\epsilon\to 0}G(C_1,C_2)=\biggl(mC_1^2/2-f_oC_2\biggr)_{\epsilon=0}$.\break So, one can
choose this functionality for $\epsilon\not=0$ and has the constant of motion given by
$$K={m\over 2}[v-h_1(t)]^2-f_o[x-vt+th_1(t)-h_2(t)]\ ,\eqno(19)$$
where $h_1$ and $h_2$ have been defined as
$$h_1(t)={1\over m}\int f(t)~dt\ ,\eqno(20a)$$
and
$$h_2(t)=\int h_1(t)~dt\ .\eqno(20b)$$
The expression (19) can also be written as
$$K=K_o(x,v) +V_{\epsilon}(v,t)\ ,\eqno(21a)$$
where $K_o$ and $V_{\epsilon}$ have been defined as
$$K_o(x,v)={1\over 2}mv^2-f_ox\ ,\eqno(21b)$$
and
$$V_{\epsilon}=-mvh_1(t)+f_ovt+{1\over 2}mh_1^2(t)-f_oth_1(t)+f_oh_2(t)\ .\eqno(21c)$$
One can see that the following limit is satisfied
$$\lim_{\epsilon\to 0}V_{\epsilon}(v,t)=0\ .\eqno(21d)$$
In particular, for a periodic function,
$$g(t)=\sin(\Omega t)\ ,\eqno(22)$$
one gets
$$K=K_o+{\epsilon f_ov\over\Omega}\cos(\Omega t)+{f_o^2\epsilon^2\over
2m\Omega^2}\cos^2(\Omega t)-{\epsilon f_o^2\over m\omega^2}\sin(\Omega t)\ .\eqno(23)$$
Since the expressions (14a) and (14b) represent a no-autonomous system, the possible
associated Hamiltonian can not be a constant of motion, and the expression (7) can not be used
[7] to calculated the Lagrangian of the system, therefore its Hamiltonian. Naively, one can
consider (14a) and (14b) as a Hamiltonian system and to get $H=p^2/2m-f(t) x/m$ as its
associated Hamiltonian ($p=mv$), and $L=mv^2/2+f(t) x/m$ as its associated Lagrangian. However,
this procedure is hardly to justify, and it is not free of ambiguities.
\vskip1pc
\leftline{\bf III. Constant of motion of a time depending mass system.}
\vskip0.5pc
The motion of a time depending mass under a constant force can be described by the following
no-autonomous dynamical system
$${dx\over dt}=v\eqno(24a)$$
and
$${dv\over dt}={f\over m}-{\dot m\over m}v\ ,\eqno(24b)$$
where $f$ represents the constant force, $m=m(t)$ is the  mass of the system, and $\dot m$ is
its time differentiation. A constant of motion for this system is a function $K(x,v,t)$
which satisfies the equation
$$v{\partial K\over\partial x}+\left[{f\over m}-{\dot m\over m}v\right]{\partial
K\over\partial v}+{\partial K\over\partial t}=0\ .\eqno(25)$$
Solving (25) by the characteristics method, the general solution is gotten as
$$K(x,v,t)=G(C_1,C_2)\ ,\eqno(26)$$
where $G$ is an arbitrary function of the characteristics $C_1$ and $C_2$ which are defined as
$$C_1= mv-ft\eqno(27a)$$
and
$$C_2=x-mv\int{dt\over m(t)}+f\left[t\int{dt\over m(t)}-\int {t~dt\over m(t)}\right]\
.\eqno(27b)$$  
If one assumes that the mass is constant, $m(t)=m_o$, the characteristics curves would be
given by $C_1=mv-ft$ and $C_2=x-vt-ft^2/2m$. So, the functionality $G$ which brings
about the usual constant of motion (energy)  would be given by $G=C_1^2/2m_o-fC_2=mv^2/2-fx$.
Therefore, for the case where the mass depends explicitly on time and of the form
$$m(t)=m_og_{\epsilon}(t)\eqno(28)$$
such that $\lim_{\epsilon\to 0}g_{\epsilon}=1$, one chooses
$$G(C_1,C_2)={1\over 2m_o}C_1^2-fC_2\eqno(29)$$
which brings about the constant of motion of the form
$$K_{\epsilon}(x,v,t)=K_{o\epsilon}(x,v,t)+ W_{\epsilon}(v,t)\ ,\eqno(30a)$$
where $K_{o\epsilon}$ and $W_{\epsilon}$ are given by
$$K_{o\epsilon}={m_og_{\epsilon}^2\over 2}v^2-fx\eqno(30b)$$
and
$$W_{\epsilon}=-g_{\epsilon}(t)fvt+{f^2t^2\over 2m_o}+g_{\epsilon}(t)fv\Lambda_1(t)-
{f^2\over m_o}\Lambda_2(t)\ .\eqno(30c)$$
The functions $\Lambda_1(t)$ and $\Lambda_2(t)$ have been defined as
$$\Lambda_1(t)=\int{dt\over g_{\epsilon}(t)}\eqno(30d)$$
and
$$\Lambda_2(t)=t\Lambda_1(t)-\int{t~dt\over g_{\epsilon}(t)}\ .\eqno(30e)$$
The functions $K_{o\epsilon}$ and $W_{\epsilon}$ have the following limits
$$\lim_{\epsilon\to 0}K_{o\epsilon}={1\over 2}m_ov^2-fx\eqno(31a)$$
and
$$\lim_{\epsilon\to 0}W_{\epsilon}=0\ .\eqno(31b)$$
The observation about getting the Hamiltonian for this system, equations (24a) and (24b), is
essentially the same as previous system, and it will not be discussed any further.
\vfil\eject
\leftline{\bf IV. Constant of motion of a position depending mass system.}
\vskip0.5pc
The motion of a relativistic particle of position depending mass, $m(x)$, under a conservative
force $f(x)$ is given by the equation
$${d\over dt}\left({m(x)v\over\sqrt{1-{v^2\over c^2}}}\right)=f(x)\ ,\eqno(32)$$
where $v$ is the velocity of the particle. This equation can be written as the following
autonomous system
$${dx\over dt}=v\eqno(33a)$$
and
$${dv\over dt}={f(x)\over m}\left(1-{v^2\over c^2}\right)^{3/2}-
\left(1-{v^2\over c^2}\right){v^2m_x\over m}\ ,\eqno(33b)$$
where $m_x$ is the differentiation of the mass $m$ with respect the position. A constant of
motion for this system is a function $K(x,v)$ satisfying the equation
$$v{\partial K\over\partial x}+
\left[\left(1-{v^2\over c^2}\right)^{3/2}{f(x)\over m}-\left(1-{v^2\over
c^2}\right){v^2m_x\over m}\right]{\partial K\over\partial v}=0\ .\eqno(34)$$
The general solution of (34) is given by 
$$K(x,v)=G(C)\ ,\eqno(35)$$
where $C$ is the characteristic curve obtained from the solution of
$${dx\over v}={dv\over\ds \left(1-{v^2\over c^2}\right)^{3/2}{f(x)\over m}-\left(1-{v^2\over
c^2}\right){v^2m_x\over m} }\ .\eqno(36)$$
From this expression, one can see clearly that this equation can be integrated for special 
cases only. For example, assuming $f(x)$ of the form
$$f(x)=-\alpha m_xc^2\ ,\eqno(37)$$
where $\alpha$ is a constant. Using (37) in (36) and a new variable $\xi=\sqrt{1-v^2/c^2}$,
the integration can be done, getting the characteristic curve (in terms of the variable $v$)
$$C_{\alpha}=m\sqrt{v^2/c^2+\alpha\sqrt{1-v^2/c^2}\over 1-v^2/c^2}
\left({\sqrt{\alpha^2+4}-\alpha+2\sqrt{1-v^2/c^2}\over
\sqrt{\alpha^2+4}+\alpha-2\sqrt{1-v^2/c^2}}\right)^{\ds\alpha\over\ds 2\sqrt{\alpha^2+4}}\
.\eqno(38)$$
Note, from (7), that $\alpha=0$ represents the case of a relativistic free particle with
position depending mass, and from (38) one gets the following limit
$$\lim_{\alpha\to 0}C_{\alpha}={m(x)\over c\sqrt{1-v^2/c^2}}\ .\eqno(39)$$
Thus, one can choose $G$ of the form $G(C_{\alpha})=c^2 C_{\alpha}^2/2m_o$, where $m_o$
is the value of $m$ at $x=0$,  to get the constant of motion
$$K_{\alpha}=\left({m^2(x)\over 2m_o}\right){v^2+\alpha c^2\sqrt{1-v^2/c^2}\over 1-v^2/c^2}
 \left({\sqrt{\alpha^2+4}-\alpha+2\sqrt{1-v^2/c^2}\over
\sqrt{\alpha^2+4}+\alpha-2\sqrt{1-v^2/c^2}}\right)^{\ds\alpha\over\ds \sqrt{\alpha^2+4}}\
.\eqno(40)$$
In addition, if $m(x)$ is of the form
$$m(x)=m_og_{\epsilon}(x)\ ,\eqno(42)$$
where $\lim_{\epsilon\to 0}g_{\epsilon}(x)=1$, one would have the following expected limit
$$\lim_{{\scriptstyle\alpha\to 0\atop\scriptstyle c\to\infty}\atop\scriptstyle \epsilon\to
0}K_{\alpha}={1\over 2}m_ov^2\ .\eqno(43)$$
For example, choosing $m(x)$ as
$$m(x)=m_o\bigl(1+\epsilon\sin(kx)\bigr)\ ,\eqno(44)$$
the constant of motion is written as
$$K_{\alpha}(x,v)={m_oc^2\over 2}\biggl(1+\epsilon\sin(kx)\biggr)^2F_{\alpha}\left({v\over
c}\right)\ ,\eqno(45)$$
where the function $F_{\alpha}$ si given by
$$F_{\alpha}\left({v\over c}\right)={v^2/c^2+\alpha\sqrt{1-v^2/c^2}\over 1-v^2/c^2}
\left({\sqrt{\alpha^2+4}-\alpha+2\sqrt{1-v^2/c^2}\over
\sqrt{\alpha^2+4}+\alpha-2\sqrt{1-v^2/c^2}}\right)^{\ds\alpha\over\ds\sqrt{\alpha^2+4}}\
.\eqno(46)$$ Given the initial condition ($x_o, v_o$), this constant is determined, and the
trajectories in the space ($x, v$) can be traced. On the other hand, for this system,
equations (33a) and (33b), and for the particular case seen above which brings about the
constant of motion (40), the expression (7) can be used, in principle, to obtain the
Lagrangian of the system. However, the integration can not be done in general. Even more,
if this Lagrangian is explicitly known and the generalized linear momentum is calculated, one
can not know $v=v(x,p)$, in general. Thus, the Hamiltonian of the system can not known
explicitly.  
\vfil\eject
\leftline{\bf V. Conclusions.}
\vskip0.5pc\noindent
We have given the constant of motion for several one-dimensional systems. These constants of
motion were chosen such that  they can have the usual  energy expression  when the
parameter which characterizes the no-conservative interaction goes to zero. For a relativistic
particle with linear dissipation, its constant of motion was deduced in general, but its
Hamiltonian was explicitly given only for weak dissipation. For a no-relativistic time
depending system, for a no-relativistic time depending mass system under a constant force, and
for a mass position depending system under a constant force,  only the constants of motion were
given, outlining the problem of getting their Hamiltonians. 
\vfil\eject
\leftline{\bf APPENDIX}
\vskip1pc\noindent
The function $G_{\beta}(v)$ is given by
$$\eqalign{
G_{\beta}(v)=&-2\sqrt{2+2\beta c}~v\arctan\left({v\sqrt{1+\beta c}\over c\sqrt{2}}\right)+
4\beta cv\ln\left({v/c\over \beta v-1}\right)\cr
&-4c\ln\left({2\beta c\bigl(-1+\beta c+\sqrt{\beta^2c^2-1}\sqrt{1-v^2/c^2}\over
1-\beta v}\right)\cr
&-\sqrt{2+2\beta c}~vH_{\beta}^+(v)+\sqrt{2+2\beta c}~vh_{\beta}^{-}(v)\ ,\cr}\eqno(A_1)$$
where the function $h_{\beta}^s$ with $s=\pm 1$ is given by
$\bigl(\gamma^{-1}=\sqrt{1-v^2/c^2}~\bigr)$
$$h_{\beta}^s(v)=\ln\left[{4v\sqrt{(\beta c)^2-1}+s2\beta c^2\sqrt{2+2\beta
c}\gamma^{-1}+2c\sqrt{2+2\beta c}\bigl(s\sqrt{(\beta c)^2-1}-
s\gamma^{-1}\bigr)\over
(\beta c-1)\sqrt{\beta^2c^2-1}\bigl(sc\sqrt{2+2\beta c}+v+\beta cv\bigr)}\right]\ .\eqno(A_2)$$
The function $Q_{\beta}(v)$ is given by
$$\eqalign{
Q_{\beta}(v)=&\sqrt{1-(\beta c)^2}\arctan\left({\beta c-v/c\over\sqrt{1-(\beta
c)^2}\gamma^{-1}}\right)\cr
&+\beta cv\sqrt{1-(\beta c)^2}\ln\left({2\left({\beta c-v/c\over 
1-(\beta c)^2}+{1\over\gamma\sqrt{1-(\beta c)^2}}\right)\over 1-\beta v}\right)\cr
&+(1-\beta^2c^2)\ln\left({2c\bigl(1-\beta^2c^2+(1-\beta^2c^2)\sqrt{1-v^2/c^2}\over
v(1-\beta^2c^2)^{3/2}}\right)\ .\cr}\eqno(A_3)$$
The function $R_{\beta}(v)$ is given by
$$R_{\beta}(v)=\ln\left({2c\bigl(1+\gamma^{-1}\bigr)\over v}\right)\ .\eqno(A_4)$$
The function $A_{\beta}^{(1)}$ is given by
$$\eqalign{
A_{\beta}^{(1)}(v)&={mc^3v\over \beta^2c^2-1}\Biggl\{{\beta\over v}-{\beta^2\over \beta v-1}-
{1+\beta c\over 2c^2\left(1-{(1+\beta c)v^2\over 2c^2}\right)}\cr
&+{(\beta^2c^2-1)f_1(v)\over g_1(v)\bigl(-\sqrt{2}~c\sqrt{1+\beta c}+v+\beta c v\bigr)^2}\cr
&-{(\beta^2c^2-1)f_2(v)\over g_2(v)\bigl(\sqrt{2}~c\sqrt{1+\beta c}+v+\beta c v\bigr)^2}\cr
&-{f_3(v)\over g_3(v)(1-\beta v)^2}\cr
&+{1\over v^2}\ln\left({2(-\beta c+\beta^2c^2)+2\beta c\sqrt{(\beta c)^2-1}~\gamma^{-1}\over
1-\beta v}\right)~\Biggr\}\cr}\eqno(B_1)$$
where $f_1$, $f_2$, $f_3$, $g_1$, $g_2$ and $g_3$ are defined as
$$\eqalign{
f_1(v)&={2(1+\beta c)\bigl(\sqrt{2}~c\sqrt{1+\beta c}-2 v\bigr)\over 2(\beta c-1)}
+{4\over \beta c-1}\bigl(-\sqrt{2} c\sqrt{1+\beta c}+v+\beta cv\bigr)\cr
&+{2\sqrt{2}\sqrt{1+\beta c}~v\bigl(-\sqrt{2}~c\sqrt{1+\beta c}+v+\beta cv\bigr)\over
c\sqrt{(\beta c)^2-1}~\gamma^{-1}}
+{2\sqrt{2} c(1+\beta c)^{3/2}~\gamma^{-1}\over \sqrt{\beta^2c^2-1}}\ ,\cr}\eqno(b1)$$
$$\eqalign{
f_2(v)&=-{2(1+\beta c)\bigl(\sqrt{2}~c\sqrt{1+\beta c}+2 v\bigr)\over 2(\beta c-1)}
+{4\over \beta c-1}\bigl(\sqrt{2} c\sqrt{1+\beta c}+v+\beta cv\bigr)\cr
&-{2\sqrt{2}\sqrt{1+\beta c}~v\bigl(-\sqrt{2}~c\sqrt{1+\beta c}+v+\beta cv\bigr)\over
c\sqrt{(\beta c)^2-1}~\gamma^{-1}}
-{2\sqrt{2} c(1+\beta c)^{3/2}~\gamma^{-1}\over \sqrt{\beta^2c^2-1}}\ ,\cr}\eqno(b2)$$
$$f_3(v)=
-{2\beta\gamma v\over c}\sqrt{\beta^2c^2-1}~(1-\beta v)+
\beta\bigl[2(-\beta c+\beta^2c^2)+2\beta c\sqrt{(\beta c)^2-1}~\gamma^{_1}\ ,\eqno(b3)$$
$$\eqalign{
&g_1(v)=2\sqrt{2}~c(\beta c-1)\sqrt{1+\beta c}\cr
&\times\left[-{2\bigl(\sqrt{2} c\sqrt{1+\beta c}-2v\bigr)\over
(\beta c-1)\bigl(-\sqrt{2}~c\sqrt{1+\beta c}+v+\beta vc\bigr)}
-{2\sqrt{2}~c\sqrt{1+\beta c}~\gamma^{-1}\over
\sqrt{(\beta c)^2-1}\bigl(-\sqrt{2}~c\sqrt{1+\beta c}+v+\beta cv\bigr)}\right]\
,\cr}\eqno(b4)$$
$$\eqalign{
&g_2(v)=2\sqrt{2}~c(\beta c-1)\sqrt{1+\beta c}\cr
&\times\left[{2\bigl(\sqrt{2} c\sqrt{1+\beta c}-2v\bigr)\over
(\beta c-1)\bigl(\sqrt{2}~c\sqrt{1+\beta c}+v+\beta vc\bigr)}
+{2\sqrt{2}~c\sqrt{1+\beta c}~\gamma^{-1}\over
\sqrt{(\beta c)^2-1}\bigl(\sqrt{2}~c\sqrt{1+\beta c}+v+\beta cv\bigr)}\right]\
,\cr}\eqno(b5)$$
and 
$$g_3(v)=v\bigl[2(-\beta c+\beta^2c^2)+2\beta c\sqrt{(\beta c)^2-1}~\gamma^{-1}\bigr]\
.\eqno(b6)$$
The function $A_{\beta}^{(2)}(v)$ is given by
$$\eqalign{
A_{\beta}^{(2)}(v)&={mc^2\beta\over(1-\beta^2c^2)^{3/2}}\Biggl[
\beta c\ln\left({{2(\beta c-v/c)\over 1-\beta^2c^2}+{2\over\gamma\sqrt{1-\beta^2c^2}}\over
1-\beta v}\right)\cr\cr
&-\sqrt{1-\beta^2c^2}\ln\left({2c\bigl(1-\beta^2c^2+(1-\beta^2c^2)\gamma^{-1}\over
v(1-\beta^2c^2)^{3/2}}\right)\Biggr]\ .\cr}\eqno(B_2)$$

\vfil\eject
\leftline{\bf References}
\vskip0.5pc
\obeylines{
[1] S. Okubo, Phys. Rev. A,23 (1981)2776.
\quad F. Cantrijn, J. Math. Phy. 23 (1982)1589.
\quad G. L\'opez, M. Murgu\'{\i}a and M. Sosa, Mod. Phy. Lett. B, 15, 22(2001)965.
[2] G. L\'opez, Rev. Mex. Fis., 48 (2002)10.
\quad G. L\'opez, Int. Jou. Theo. Phy. 37,5 (1998)1617.
[3] G. L\'opez, Rev. Mex. Fis., 45,6 (1999)551.
[4] G. L\'opez, "Partial Differential Equations of First Order and Their Applications
\quad to Physics," World Scientific, 1999.
[5] J.A. Kobussen, Acta Phy. Austriaca, 51 (1979)293.
\quad C. Leubner, Physica A 86 (1981)2.
\quad G. L\'opez, Ann. of Phy., 251,2 (1996)372.
[6] F. John, "Partial Differential Equations," Springer-Verlag, N.Y. 1974.
[7] G. L\'opez and J.I. Hern\'andez, Ann. of Phy., 93,1 (1989)1.
}

\end